\documentclass[prl,twocolumn,showpacs,amsmath,amssymb,superscriptaddress]{revtex4-1}

\usepackage{graphicx}% Include figure files
\usepackage{dcolumn}% Align table columns on decimal point
\usepackage{bm}     % bold math
\usepackage{bbm}
\usepackage{multirow}
\usepackage{booktabs}
\usepackage{amsmath}
\usepackage{braket}
\usepackage{appendix}
\usepackage{float}
\usepackage{xcolor}
\usepackage{physics}
\usepackage{siunitx}
\usepackage{amssymb}
\usepackage{array}
\usepackage{hyperref}

\begin{document}
	\title{Thermal melting of a quantum electron solid in the presence of strong disorder: Anderson versus Wigner}
	
	\author{DinhDuy Vu}
	\author{Sankar Das Sarma}
	\affiliation{Condensed Matter Theory Center and Joint Quantum Institute, Department of Physics, University of Maryland, College Park, Maryland 20742, USA}
	
	\begin{abstract}
		We consider temperature-induced melting of a Wigner solid in one dimensional (1D) and two dimensional (2D) lattices of electrons interacting via the long-range Coulomb interaction in the presence of strong disorder arising from charged impurities in the system.  The system simulates semiconductor-based 2D electron layers where Wigner crystallization is often claimed to be observed experimentally.  Using exact diagonalization and utilizing the inverse participation ratio as well as conductance to distinguish between the localized insulating solid phase and the extended metallic liquid phase, we find that the effective melting temperature may be strongly enhanced by disorder since the disordered crystal typically could be in a localized glassy state incorporating the combined nonperturbative physics of both Anderson localization and Wigner crystallization.  This disorder-induced enhancement of the melting temperature may explain why experiments often manage to observe insulating disorder-pinned Wigner solids in spite of the experimental temperature being decisively far above the theoretical melting temperature of the pristine Wigner crystal phase in many cases.
	\end{abstract}
	\maketitle
	
	\textit{Introduction and background - }
	Wigner predicted more than 80 years ago that an interacting electron liquid  would crystallize at $T=0$ into a quantum solid at a sufficiently low density, below some dimension-dependent critical density, because of their long range Coulomb interaction \cite{Wigner}. The basic idea is simple:  the quantum non-interacting kinetic energy  (i.e. “quantum fluctuations”) for an electron density of $n$ in a $d$-dimensional system typically goes as $k_F^2\sim n^{-2/d}$ (for parabolic band dispersion typical of semiconductors) simply by virtue of the uncertainty principle whereas the Coulomb interaction potential energy ($\sim$ 1/distance) goes as $n^{-1/d}$, implying that for low enough $n$, the system would minimize the potential energy by crystalizing into a solid phase instead of being a liquid (as happens at higher densities) which minimizes the kinetic energy. Typically, the critical density separating the quantum Wigner solid and the electron liquid phase is very low, and observing a density-tuned quantum Wigner solid is a huge experimental challenge.  On the other hand, if the temperature is much higher than the non-interacting Fermi temperature, the electrons undergo a liquid-to-solid transition with increasing density (in contrast to the quantum crystallization, happening with decreasing density at $T=0$) at a finite temperature. This is because the delocalization effect is now generated by thermal fluctuations $\propto T\sim n^0$ instead of quantum fluctuations $\sim n^{-2/d}$. Such a classical liquid to solid  transition was indeed observed in two-dimensional (2D) electrons on the surface of liquid helium a long time ago \cite{Grimes1979, Marty1980}

	The current work focuses on the thermal melting of the quantum Wigner solid ($T<T_F$) in the presence of disorder.  The density-temperature mean field phase diagram of the disorder-free pristine electron solid to liquid transition has been calculated in earlier works \cite{Hwang2001,Vu2020}. In 2D systems, the solid-to-liquid density-tuned $T=0$ quantum transition occurs at $r_s=r_c \sim 40$, where $r_s$ is the usual dimensionless average electron separation measured in the units of effective Bohr radius with large $r_s$ implying strong interaction.  For $r_s=r_c$, the melting temperature of the Wigner solid is $T_m=0$, and $T_m$ increases with increasing $r_s$, eventually at large enough $r_s$, at a given $T$, the system enters the classical regime where $T>T_F$. As a demonstration, for GaAs,  the highest melting temperature for the Wigner solid was determined to be $\sim 10$ mK for $r_s\sim60 \sim 1.5 r_c$ \cite{Hwang2001} by  interpolating between the classical \cite{Gann1979} and the quantum \cite{Tanatar1989, Drummond2009} theories. It is therefore extremely puzzling that there are many experimental claims of the observation of 2D quantum Wigner crystallization in GaAs based on transport measurements in 2D electron (or hole) gases confined in semiconductor layers, typically carried out around $r_s \sim 25-40$, where the theoretically predicted $T_m$ is much smaller than the experimental base temperature of 20 mK in the dilution fridge  \cite{Shayegan,Goldman1990,Yoon1999}. $T_m$ scales approximately as $\sim m/\kappa^2$, where $m, \kappa$ are the effective mass and the background dielectric constant, so the maximum Wigner crystal melting temperature can reach  $70~$mK for ZnO,  $90~$mK for AlAs quantum well and as high as $1~$K for MoSe\textsubscript{2} monolayers. In these new platforms, the critical $r_s\sim 30$ also corresponds to a higher value of electron density, making them better candidates to host Wigner crystal. In fact, the Wigner crystal has been claimed to be observed in these systems \cite{Hossain2020,smolenski2021,Falson2022,Hossain2022}. However, our goal is not disproving the Wigner crystal claim in all experiments simply because the melting temperature is higher than the theoretically predicted value, but instead pointing out a real possibility that if the impurity density is comparable to the carrier density, the disorder-induced insulator can exhibit the same localized-delocalized transition as the thermal melting of the Wigner crystal with increasing temperature.

	In the current work, we propose and validate by exact small system numerical calculations that the correct interpretation of these GaAs-based Wigner crystal experimental observations lies in including disorder effects in the theory. In particular, disorder in some sense enhances the effective melting temperature of the localized solid insulating phase compared with the extremely low melting temperature of the pristine system. Since the disorder in the high quality semiconductor layers arises invariably from random charged impurities which also interact with the electrons via the same long range Coulomb coupling as the electron-electron mutual interaction itself is, it is imperative to include both electron-impurity and electron-electron interactions on an equal footing.  In fact, the $r_s$ value of 40 ($\sim r_c$) in 2D GaAs electron layers corresponds to a 2D carrier density of $< 10^9 \text{cm}^{-2}$, which is likely to be comparable to or smaller than the random charged impurity density even in the highest quality semiconductor structures \cite{DasSarma2015,Watson2011}, thus reinforcing the necessity of treating both electron-electron and electron-impurity interactions on an equal nonperturbative footing. 
	
	\textit{Model and theory - }
	In order to calculate the transport properties of an interacting electron system nonperturbatively in the presence of quenched charged impurities, we set the impurity potential is thus deterministic instead of the usual random disorder model. We perform simulations on both the 1D and 2D geometries. Since our interest is the thermal melting of the electron solid phase, the 1D calculation should provide a good qualitative description for what happens also in 2D because the pristine Wigner solid to quantum liquid density-temperature phase diagrams are essentially identical qualitatively in 1D and 2D as can be seen by comparing Fig. 8 in Ref.~\cite{Vu2020} for the 1D phase diagram with Fig. 4 in Ref.~\cite{Hwang2001} for the 2D phase diagram. Most of this paper presents 1D results but in fact, our 1D and 2D calculations produce very similar features. The small system size of our simulation makes the localized-delocalized transition necessarily a crossover, so what we address is not so much the absolute quantitative phase diagram, but how increasing disorder affects the transport properties at finite temperatures, particularly whether localization persists to higher temperatures in the solid phase with increasing disorder.
	
	Our model is minimal: A finite 1D chain (or 2D square) of $N_s =16$ lattice sites with  $N_e=4$  interacting spinless fermions  with a hopping energy $t=1/a^2$, where $a$ is the lattice constant and with periodic boundary conditions.  The use of spinless fermions is only for computational convenience, and affects no aspect of the key issues we address as we are not interested in the difficult energetic question of whether the solid phase is ferromagnetic or not (which necessitates a calculation of very small exchange energies to better than $10^{-6}$ precision) \cite{Vu2020}. The 1D interacting Hamiltonian (see the Supplemental Materials~\cite{Supplement} for the 2D Hamiltonian) is given by
	\begin{equation}
	H = \sum_i t(c_i^\dagger c_{i+1}+c_{i+1}^\dagger c_i) + V_in_i +\sum_{i<j} U_{i,j}n_in_j. 
	\end{equation}
	with the interaction $U$ being the long-range Coulomb coupling
	\begin{equation}
	U_{i,j} = \left[\frac{aN_s}{\pi} \left|\sin \frac{\pi(i-j)}{N_s} \right|\right]^{-1}
	\end{equation}	
	and the impurity-induced potential $V$ being
	\begin{equation}
	V_i = -\frac{V}{a}\sum_j\left[\frac{N_s}{\pi} \left|\sin \frac{\pi(i-X_j)}{N_s} \right|+\delta\right]^{-1}
	\end{equation}
	with $\delta=1$ is the short-range regularization and $X_j$ is the position of the impurity (note that the impurity potential is long-ranged consistent with the known disorder in semiconductor materials arising from quenched charged impurities \cite{DasSarma2015,Watson2011}). The sine function ensures the periodic boundary condition. We diagonalize the Hamiltonian exactly and obtain the occupancies at finite temperature $T$ through the usual canonical thermal distribution $\langle n_i\rangle_T = \mathcal{Z}^{-1}\sum_j \bra{\psi_j}n_i\ket{\psi_j}e^{-E_j/T}$ with $\mathcal{Z}$ being the partition function.
	Using the exact $\langle n_i\rangle_T$, we calculate the key operational quantity defining transport, namely, the inverse participation ratio, IPR, defined by:
	\begin{equation}
	\text{IPR} = \left[\frac{N_s}{N_e^2}\sum_i \langle n_i\rangle_T^2\right]\left(\frac{N_s}{N_e}-1\right)^{-1}.
	\end{equation}
	It is well-known that a localized insulating (extended metallic) system has IPR = 1 (0) in the thermodynamic limit, and the IPR is used extensively in studying electronic transport properties both for interacting and for noninteracting systems.  There is, however, a serious problem in finite systems, where the IPR is always finite and $<1$ simply because  even localized states can never have strictly zero conductance in finite systems. Typically small (large) IPR values are associated with metallic (insulating) states, and we make the arbitrary distinction that IPR $>  (<)~0.4$ denotes the insulating solid (metallic liquid) phase.  This is the quantum melting analog of the extensively used Lindemann criterion in classical statistical mechanics where melting is arbitrarily defined by the thermal mean square fluctuations of the solid phase around individual equilibrium sites being larger than some predefined amount in units of inter-particle separation. We emphasize that due to the crossover characteristic, the critical IPR-value (e.g. 0.4) does not have unique value. However, we show in the Supplemental Materials that this value is reasonable across multiple system sizes; moreover, redefining the IPR criteria can change the quantitative value of the melting temperature but does not alter the picture of Wigner-Anderson competition \cite{Supplement}. One cautionary note in order to avoid any confusion is that the hopping kinetic energy $1/a^2$ in our lattice model is equivalent to the Fermi energy or density in the standard continuum electron gas model (as applies to the semiconductor-based experimental systems), so we should loosely think of the lattice spacing $a$ in our lattice model as qualitatively equivalent to $r_s$ in the electron gas model and the variation of $1/a$ as the tuning the carrier density in experiments. All the parameters in our model are dimensionless with the only relevant physical scales being the effective Bohr radius and the effective electron mass. The dimensionless $V$ is physically the impurity charge relative to the carrier charge.
	
\begin{figure}
		\includegraphics[width=0.45\textwidth]{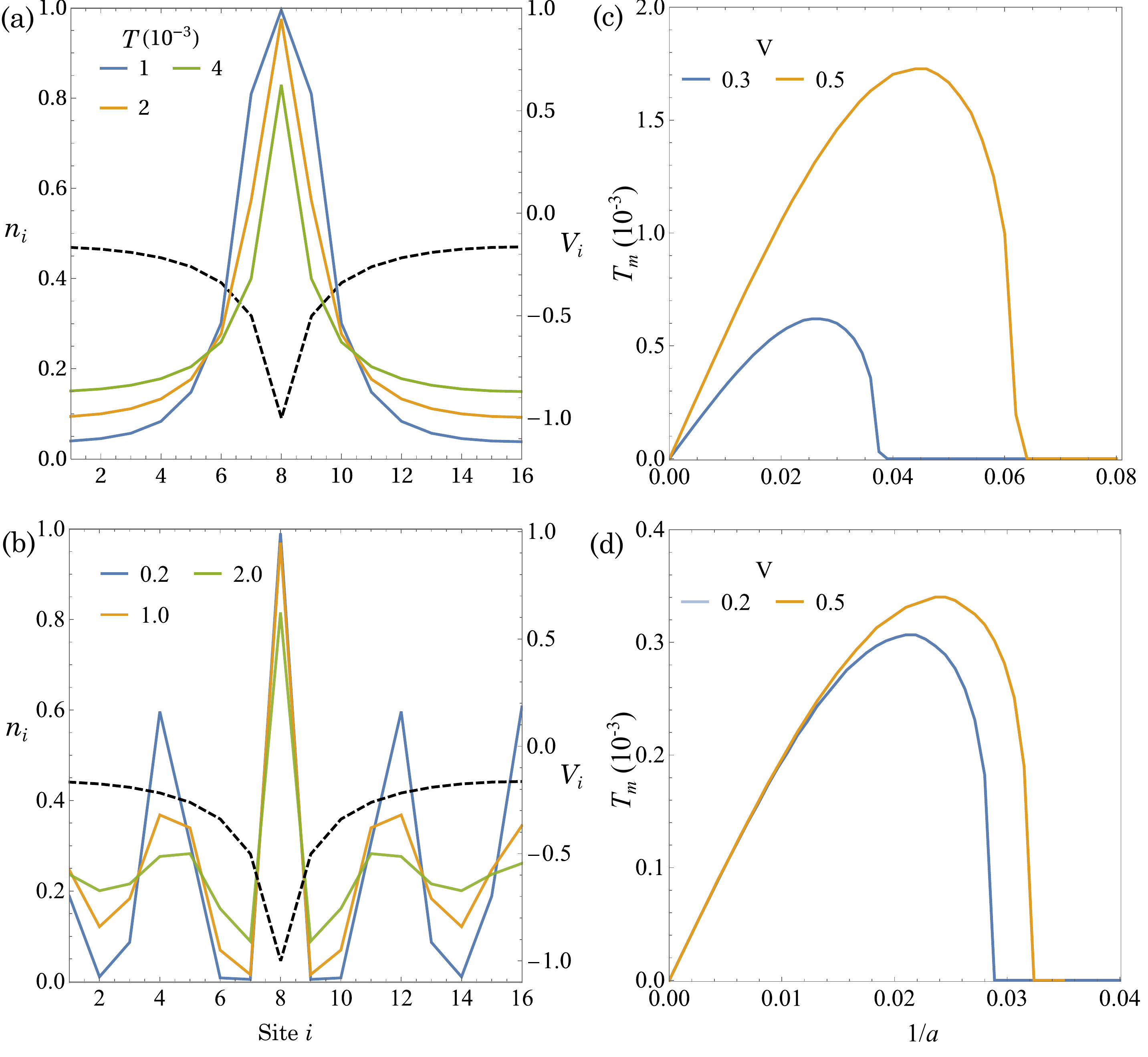}
		\caption{Density profile of the non-interacting 1D lattice with $1/a=0.03$ and $V=0.5$ (a) and the interacting 1D lattice with $1/a=0.02$ and $V=0.5$ (b). The dashed lines represent the impurity potential. (c), (d) Melting temperature of the Anderson insulator and Wigner crystal respectively.\label{fig1}}
	\end{figure}

	\textit{Results and discussion - }
	We first study the system with a single impurity. In Fig.~\ref{fig1}(a), we compute the density distribution at finite temperature of the non-interacting 1D lattice. Without the interaction, the localized phase is a disorder-induced Anderson insulator as can be seen from the bunching of electron density around the impurity potential minimum. In the presence of interaction [Fig.~\ref{fig1}(b)], the localized phase at low temperature becomes a Wigner crystal with $N_e$ equally spaced density peaks. For both Anderson insulator and Wigner crystal, increasing the temperature weakens the density modulation, resulting in a crossover to the conducting/liquid phase. 
	As mentioned above, by choosing an arbitrary critical IPR, we can identify the melting temperature separating the solid/insulating with the liquid/conducting phases. For the Wigner crystal, $T_m$ is nonmonotonic, vanishing for large and zero kinetic energy and manifesting a maxima close to the vanishing of $T_m$ at large kinetic energy. These general features also exist in the Anderson insulator because the impurity potential has the same long-range $1/r$ dependence, making the two phase diagrams (Anderson insulator and Wigner crystal) impossible to distinguish qualitatively. The distinction, however, becomes apparent when we tune the impurity strength as the phase boundary of the Anderson insulator significantly widens with increasing $V$ while that of the Wigner crystal only changes moderately mostly near the quantum critical $r_c$ [see Fig.~\ref{fig1}(c,d)]. 
	
\begin{figure}
		\includegraphics[width=0.45\textwidth]{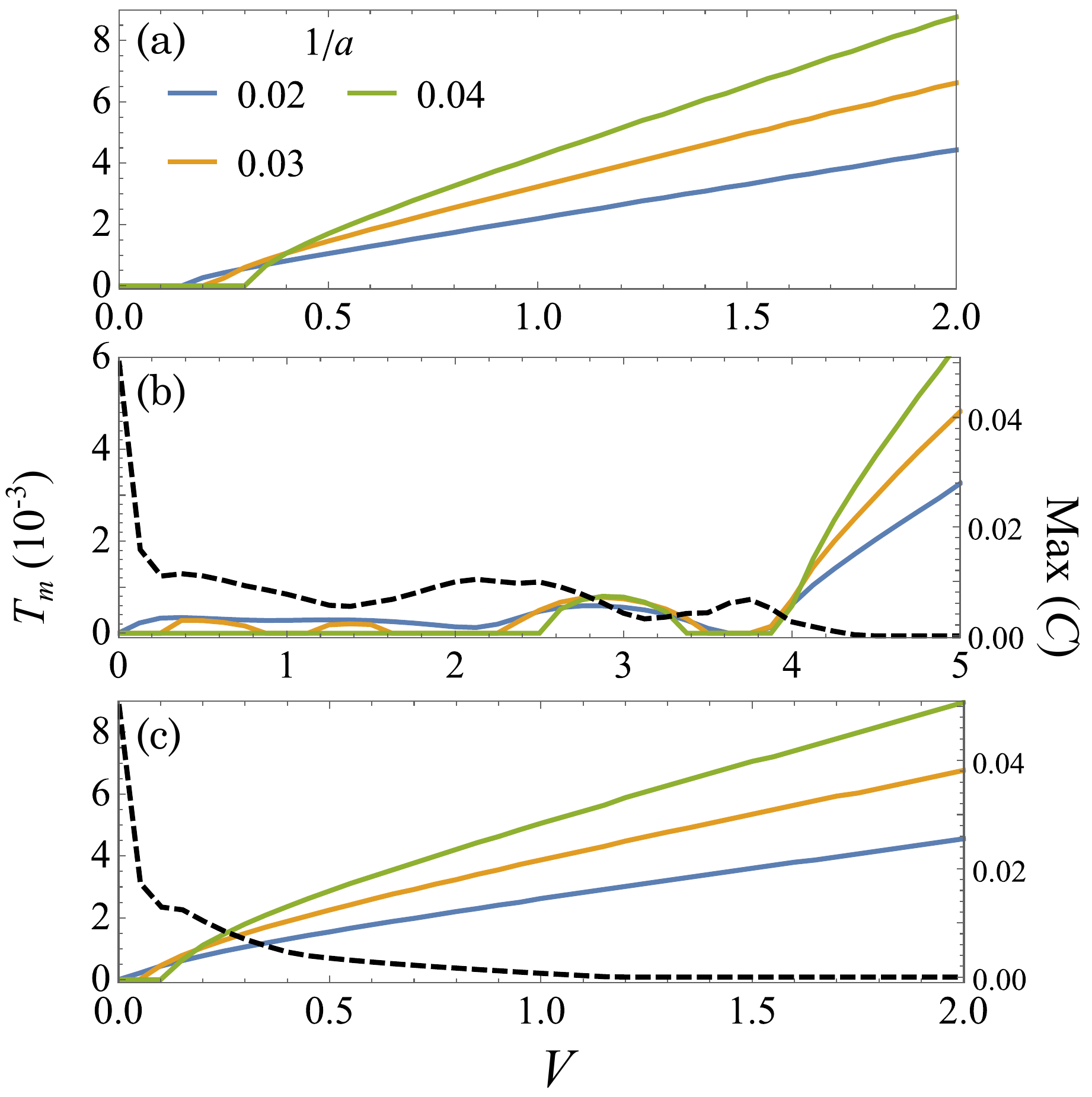}
		\caption{Melting temperature with respect to the impurity strength $V$. (a) Non-interacting 1D lattice (Anderson insulator) with $T_m$ grows linearly with $V$. (b) Interacting 1D lattice with one impurity. Depending on the impurity strength, the localized phase can be the Wigner crystal ($V<4$) or Anderson insulator phase ($V>4$). (c) Interacting 1D lattice with two impurities where the Wigner crystal phase is significantly reduced. The dashed lines are the maximum value of the density-density correlation for $1/a=0.03$ at the melting temperature for each case.\label{fig2}}
	\end{figure}

	Inspired by the preliminary results, we focus on the variation of the melting temperature with respect to the impurity strength, which is the main result of our paper. Figure~\ref{fig2}(a) shows the non-interacting (Anderson insulator) $T_m$ growing linearly with $V$ except for a very small initial range. In the linear regime, the melting temperature also progressively increases with the density, signaling the classical behavior. These feature are easy to understand if we assume the melting happens when $E_{\text{impurity}}\propto T$ with the electron-impurity energy $E_{\text{impurity}}\propto V/a$. Remarkably, in the presence of the inter-electron interaction, the similar $T_m(V)$ function displays a  Wigner-to-Anderson ‘transition’ at $V=4$ where $T_m$ increases monotonically with $V$, indicating that the insulating phase is stabilized by increasing disorder in contrast to the behavior for $V<4$ where the system is essentially a Wigner solid with small $T_m$ as found in \cite{Hwang2001,Vu2020}.  Rather amazingly, this Wigner-to-Anderson transition in Fig.~\ref{fig2}b looks essentially like a sharp transition at a critical $V (\sim 4)$ almost independent of carrier density $1/a$.  We also compute the density-density correlation
	\begin{equation}
	C(x) = \frac{1}{N_e}\sum_i \langle n_in_{i+x} \rangle_T - \langle n_i\rangle_T\langle n_{i+x}\rangle_T,
	\end{equation}
	which can only take positive values in the strongly interacting Wigner crystal phase \cite{Vu2020}. The maximum value $\text{Max}(C)$ at the melting temperature shown in Fig.~\ref{fig2}b displays a visible drop  at $V=4$, indicating that the localized phase is indeed Wigner (Anderson) for $V < (>) 4$. 
	
	We note that in our simulation, by changing the length scale $a$ while keeping the number of electrons and impurities fixed, we are effectively tuning the carrier and impurity density proportionally. In reality, only the electron density can be dynamically tuned  through the applied gate voltage. As a result, when the electron liquid is diluted experimentally to obtain the quantum Wigner crystal, its density might be comparable to (or even less than) that the impurity density. We can increase the relative $n_\text{impurity}/n_\text{electron}$ by adding more impurities to the lattice. In Fig.~\ref{fig2}c, we compute the melting temperature in the case of two impurities with equal strength $V$ placed at $X_j={1,9}$. Even though the relative impurity density only increases two fold, the critical  $V$ characterized by the sharp drop in the density correlation decreases 8-fold to $\sim 0.5$ from $\sim 4$. This shows that if the impurity density is comparable to the carrier density, the localized phase is most likely an Anderson insulator. We also repeat the computation in the case where impurities are present at all the sites but with randomized strength $\in[-V,V]$ and the critical $V$ is also significantly reduced. We believe that Figs.~\ref{fig2}(b, c) provide the explanation for why the low-density 2D electron systems generically manifest strongly insulating solid-like behavior at temperatures far above the putative melting temperature of the pristine Wigner crystal.  It is simply because the realistic solid phase is essentially an Anderson insulator with a large $T_m$ because of the invariable presence of quenched random charged impurities in the environment. 
	
	We have also calculated density patterns and the density correlations in the various finite-temperature phases ensuring that indeed the three phases, metallic liquid ($T>T_m$), Wigner solid and Anderson insulator, manifest the expected behavior well-known for these phases: metallic liquid exhibiting uniform density throughout, Wigner solid phase showing the periodic density modulation, and Anderson insulator exhibiting a density localized at impurities. We also compute the electrical conductance in the Supplemental Materials \cite{Supplement}, confirming its consistency with the IPR characterization presented in the paper.
	
	\begin{figure}
		\includegraphics[width=0.45\textwidth]{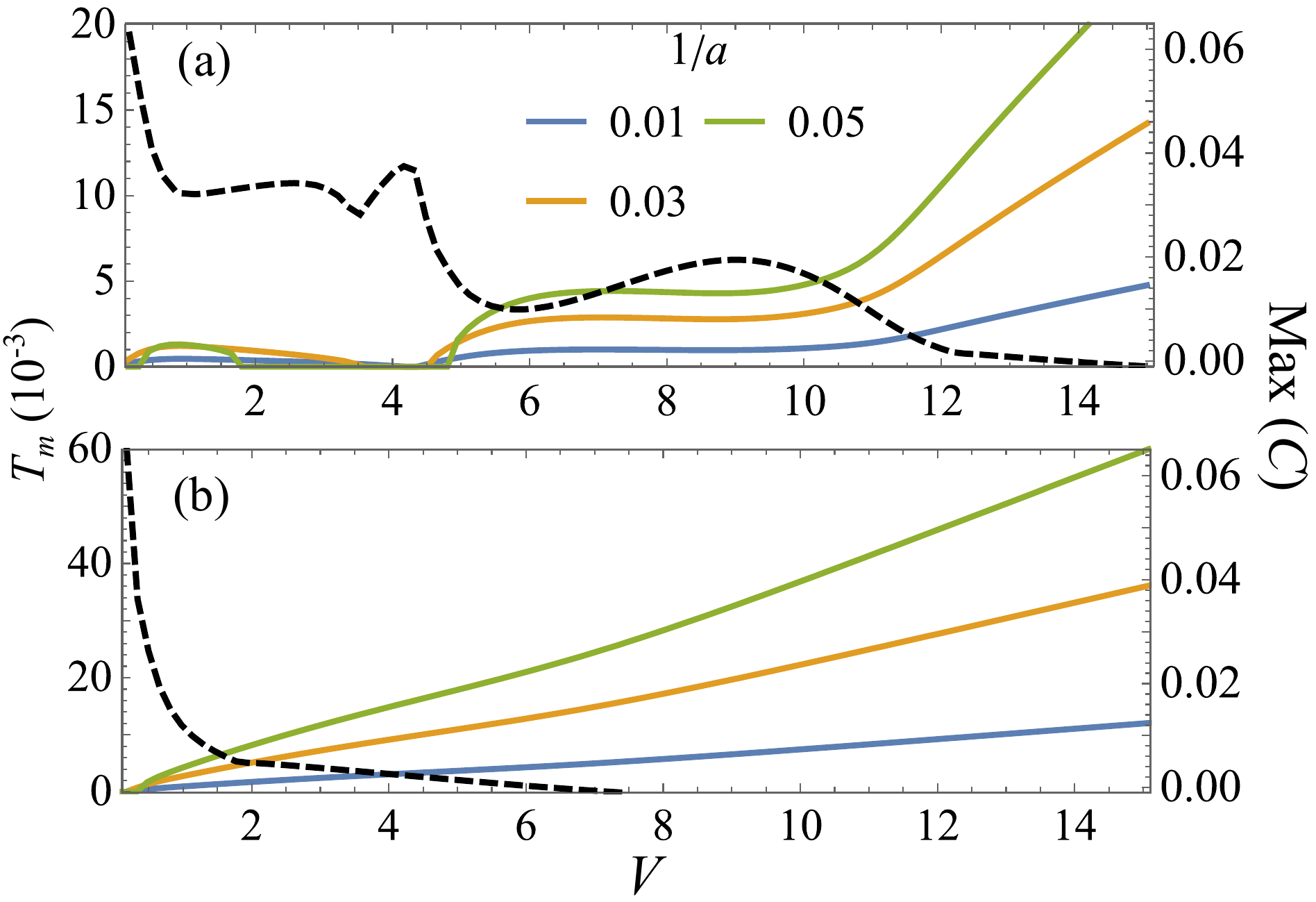}
		\caption{Same as Fig.~\ref{fig2} but for a $4\times 4$ 2D square lattice with one impurity at coordinate (1,1) (a) and two impurities at (1,1) and (2,3) (b). The lattice site label in each direction ranges from 1 to 4\label{fig3}}
	\end{figure}
	
	\textit{Relation to 2D systems -} We believe the conclusions drawn from our 1D simulations (Figs.~\ref{fig1} and \ref{fig2}) are qualitatively applicable to 2D systems as long as the electron band has quadratic dispersion and the electron-electron as well as impurity-electron interactions are long-range. Specifically, the thermal energy $\sim a^0$, the Coulomb energy $\sim a^{-1}$ and the kinetic energy $\sim a^{-2}$. The reason is that the melting of Wigner crystal or Anderson insulator is essentially due to the different scaling behaviors of the terms in the Hamiltonian, regardless of the dimensionality. Moreover, for $T=0$, both 1D and 2D systems manifest Anderson localization without a critical disorder strength. To verify this argument, we change the 16-site 1D chain into a $4\times4$ square with 2D hoppings and interactions and repeat the same IPR characterization (see \cite{Supplement} for details). The melting temperature in the case one and two impurities are shown in Figs.~\ref{fig3}(a) and (b) respectively. Similar to the 1D case, there exists a critical $V$ below (above) where the localized phase is a Wigner crystal (Anderson insulator). This critical impurity strength again (as in 1D) is significantly suppressed when the relative impurity density increases. 
	
	\textit{Conclusion - }
	Using the exact diagonalization technique, we study the thermal melting of a disordered interacting Wigner solid, finding that increasing disorder causes a sharp crossover from the Wigner solid to the Anderson insulator with a concomitant increase in the melting temperature.  A direct qualitative implication of our theory is that the experimentally observed low-density insulating phase is most likely an Anderson insulator.  Pushing the carrier density far below the critical density for the pristine electron liquid to Wigner solid phase does not favor the Wigner crystalization but in fact makes disorder effect even stronger since the impurity density in a given sample is fixed, and lowering the carrier density invariably enhances the effective disorder. The typical maximum melting temperature of the pristine Wigner solid phase is extremely low ($\sim 10$~mK for GaAs, $\sim 1~$K for TMD), but disorder enhances this melting temperature substantially creating a crossover Anderson-Wigner glassy phase with high melting temperature as shown in the current work. 
	
	The fact that disorder must always dominate the low-density phase is easily seen from a simple scaling argument involving the energetics of kinetic energy, Coulomb interaction, and Coulomb disorder, which scale respectively as $n$, $n^{1/2}$, and $n^{-1}$ as a function of the 2D carrier density $n$.  Thus, while it is indeed true that with no disorder, eventually the Coulomb interaction dominates the kinetic energy at low enough $n$, leading to Wigner crystallization at low density in the pristine system, the behavior of the disordered interacting system is fundamentally different from the pristine system since Coulomb disorder becomes by far the most dominant effect (going as $1/n$) at low enough density.  This means, no matter what, the system is asymptotically always an Anderson insulator at low enough density as long as there is any disorder!  Our exact results explicitly verify this qualitative dominance of Anderson over Wigner at low densities.  We note that the same remains true, although in a quantitatively weaker manner, for strongly screened Coulomb disorder or even strict short-range disorder since the energy scale of such a short-range disorder is $\sim n^0$ (i.e. independent of density), which would eventually dominate the kinetic energy ($\sim n$) and the Coulomb interaction energy ($\sim n^{1/2}$) at low enough density.  Thus the eventual dominance of Anderson over Wigner is guaranteed in all electronic materials studied in the laboratory, and therefore, an insulating Anderson localized phase is always the ultimate low-density phase  of all metals. An equivalent physical way of saying the same thing is that a given sample has a fixed impurity density $n_i$, and for low enough carrier density $n$, disorder must always eventually win over interaction.
	
	\begin{acknowledgments}
		\textit{Acknowledgment - }This work is supported by the Laboratory for Physical Sciences.
	\end{acknowledgments}

	\bibliographystyle{apsrev4-1}
	\bibliography{reference}
	
\end{document}

\title{Supplemental Materials for ``Thermal melting of a quantum electron solid in the presence of strong disorder: Anderson versus Wigner"}
	\maketitle
    \section{Two-dimensional lattice}
	In this section, we present the details on the 2D simulations. The Hamiltonian is given by
	\begin{equation}
		H = \frac{1}{a^2}\sum_{\langle ij \rangle} c_i^\dagger c_j + \frac{1}{a}\sum_i V_i n_i + \frac{1}{a}\sum_{i,j} U_{i,j} n_i n_j.  
	\end{equation}
    Here, $a$ is the parametric length scale (lattice constant) giving the correct scaling of each energetic term and $\langle ij \rangle$ represent nearest sites for electron hoping on a 2D geometry. Both the impurity-electron and electron-electron interactions are modeled by the long-range Coulomb potential
    \begin{equation}
    	V_i = \sum_n \frac{-V}{D(\mathbf{r}_i - \mathbf{R}_n) +\delta}, U_{i,j} = \frac{1}{D(\mathbf{r}_i-\mathbf{r}_j)},
    \end{equation}
    where the soft cutoff $\delta=1$ and the periodic 2D distance is defined as
    \begin{equation}
       D(\mathbf{r})^2 = \left( \frac{N_x}{\pi}\sin \frac{\pi x}{N_x} \right)^2 + \left( \frac{N_y}{2\pi}\sin \frac{\pi y}{N_y} \right)^2.
    \end{equation}
    Here, the coordinates of the electrons and impurities are chosen as $x_i,X_i\in\{1,\dots,N_x\}$ and $y_i,Y_i\in\{1,\dots,N_y\}$. We note that our 2D metric distance is anisotropic so that the unphysical accidental degeneracy is lifted. The Hamiltonian is solved on a PBC $4\times4$ square lattice occupied by 4 electrons. The definition of IPR and its melting criterion are similar to the 1D case, regardless of the dimensionality.
    
\begin{figure}
    	\includegraphics[width=0.48\textwidth]{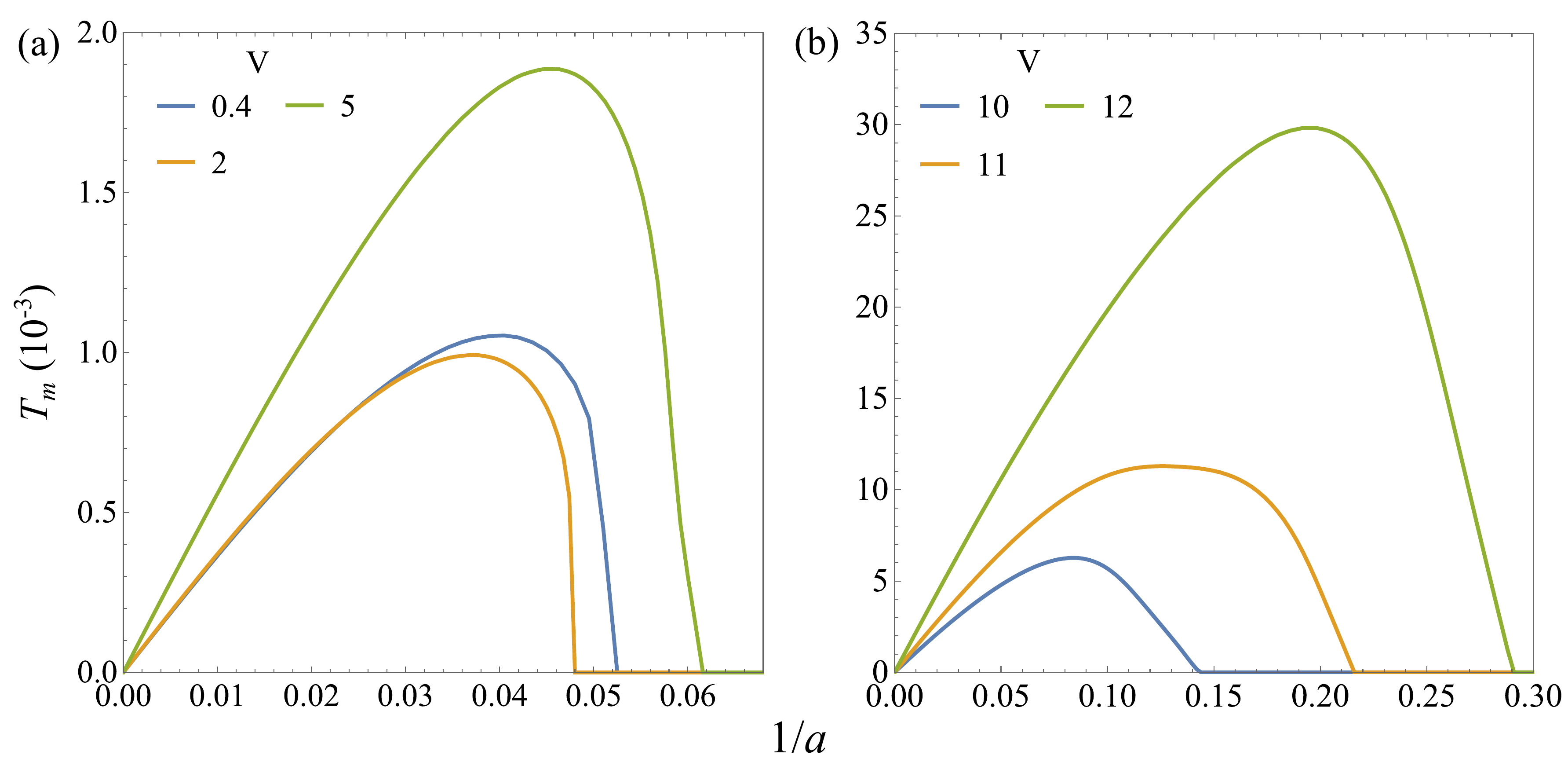}
    	\caption{Melting temperature of the Wigner crystal for $V<10$  (a) and Anderson insulator for $V\ge10$ (b) respectively.\label{fig1}}
    \end{figure}

    In Fig.~\ref{fig1}, we present the melting temperature defined as when the ground state IPR equals 0.4 as a function of the inverse length scale $1/a$ at different values of the impurity strength. The square lattice has a single impurity at $\mathbf{R}_1=(1,1)$. The phase diagram resembles the 1D case qualitatively with the linear classical branch and the almost vertical quantum branch. Consistent with Fig. 3(a) in the main text, for $V<10$ the localized phase is a Wigner crystal with the phase boundary weakly depending on $V$, while for $V>10$ the localized phase is an Anderson insulator with the insulating regime progressively expanding with increasing $V$. For any finite disorder, therefore, the system is in a crossover regime between Wigner crystal and Anderson insulator, with the Anderson insulator becoming dominant with the lowering of the carrier density (at fixed $V$) or increasing $V$ (at fixed carrier density).
    
	\section{Linear response electrical conductance}
	Using the linear response Kubo formula, the conductance can be obtained from the eigenstate spectrum by
	\begin{equation}
		\sigma_{xx}(\omega)=\frac{1-e^{-\omega/T}}{\omega\mathcal{Z}L}\sum_{m,n}e^{-\frac{E_m}{T}}\bra{m}J_x\ket{n}^2\delta\left(\frac{\omega_{mn}-\omega}{\Delta}\right),
	\end{equation}
    where $\mathcal{Z}$ is the partition function, $\omega_{mn}=|E_m-E_n|$ is difference in the eigen-energies, the frequency resolution is assumed to be $\Delta=\sqrt{t^2+T^2}$ with $t,T$ being the hoping strength and the temperature respectively, finally the scaled current operator is
    \begin{equation}
    	J_x = i\sum_{j} \left( c_{\mathbf{r}_j}^\dagger c_{\mathbf{r}_j-\mathbf{e}_x} - c_{\mathbf{r}_j}^\dagger c_{\mathbf{r}_j+\mathbf{e}_x} \right).
    \end{equation}
    Here, the unit displacement $\mathbf{e}_x = 1$ for 1D and (1,0) for 2D.
    
    In the DC limit $\omega\to 0$, the factor in front of the sum reduces to $1/T$. In addition, due to the finite resolution of the frequency, it is more meaningful to compute the scaled DC conductance $\tilde{\sigma}=T\int\sigma(\omega)F(\omega/\Delta)d\omega$ with the Lorentzian distribution $F(x)=(1+x^2)^{-1}$ (we ignore the factor $\pi$ for simplicity). As a result,
    \begin{equation}
    	\tilde{\sigma} = \frac{1}{\mathcal{Z}L}\sum_{m,n} e^{-E_m/T}\bra{m}J_x\ket{n}^2 \left(1+\frac{\omega_{mn}^2}{\Delta^2}\right)^{-1}.
    \end{equation}
   We emphasize that within our definition, the scaled  $\tilde{\sigma}$ is unchanged when the Hamiltonian and the temperature are multiplied by the same factor. The low (high) value of $\tilde{\sigma}$ is thus associated with the insulating (conducting) property of the system.

    \begin{figure}
	\includegraphics[width=0.48\textwidth]{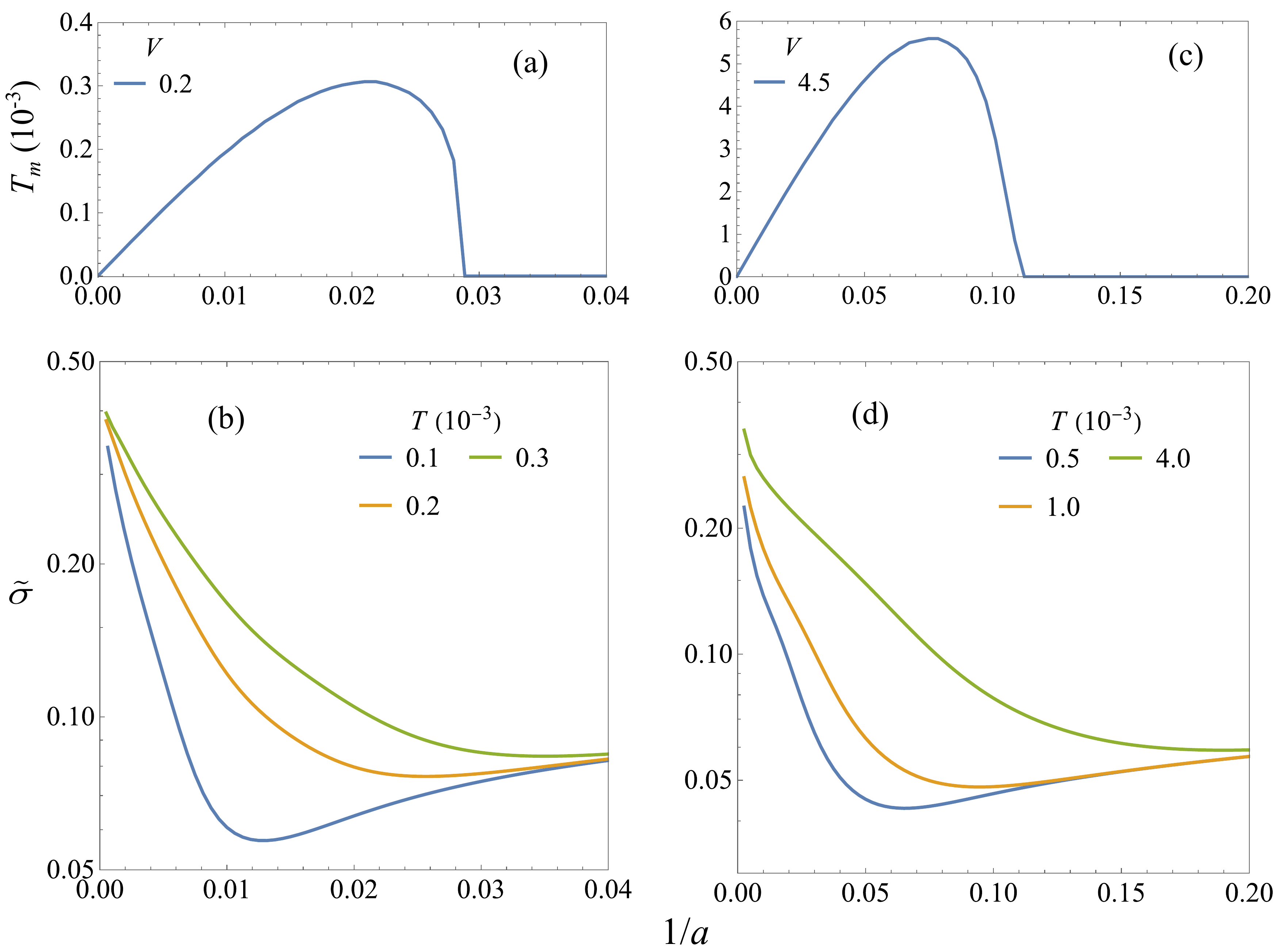}
	\caption{(a) Melting temperature of the 1D Wigner crystal  obtained from IPR. (b) Scaled conductance at the same $V$. (c-d) Same as (a-b) respectively but for the 1D Anderson insulator at large $V$.\label{fig2}}
	\end{figure}

    \begin{figure}
	\includegraphics[width=0.48\textwidth]{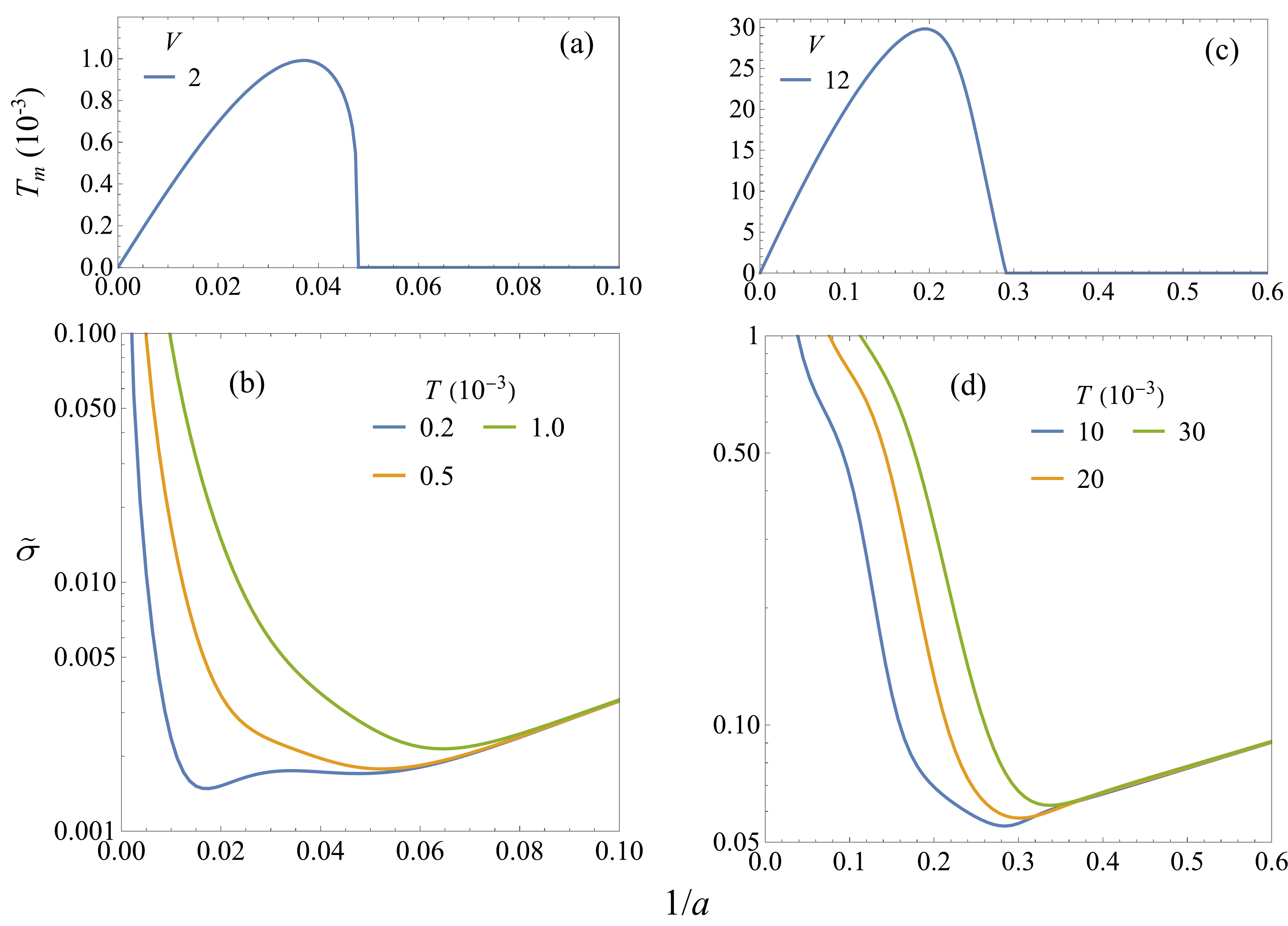}
	\caption{Same as Fig.~\ref{fig2} but for the 2D lattice.\label{fig3}}
\end{figure}
   
   In Fig.~\ref{fig2}, we present the scaled conductance of the 1D lattice with the non-monotonic behavior, i.e. $\tilde{\sigma}$ is large for low and high densities and dips at an intermediate density consistent with the localized phase characterized by the IPR. As the temperature increases, the scaled conductance increases similarly for all density, showing that the localized region shrinks at high temperature. This behavior is observed in both the Wigner crystal (low $V$) and the Anderson insulator (high $V$). We extend the conductance calculation to the 2D square lattice and obtain a qualitatively similar picture in Fig.~\ref{fig3}. We note that unlike the 1D case, in the 2D case, the conductance in Anderson localization is much higher than that of the Wigner crystal. We hypothesize that this significant disparity might be because in the Anderson insulator, the electrons bunch around the impurity along the $y-$axis, facilitating indirect hopping; while in the Wigner crystal, the electrons are spaced further apart in both directions. This difference, however is a small-system finite size effect, which should vanish when the $x$-length approaches infinity so that subleading processes are suppressed.
   \section{IPR criterion for the melting temperature}
   
   \begin{figure}
	\includegraphics[width=0.44\textwidth]{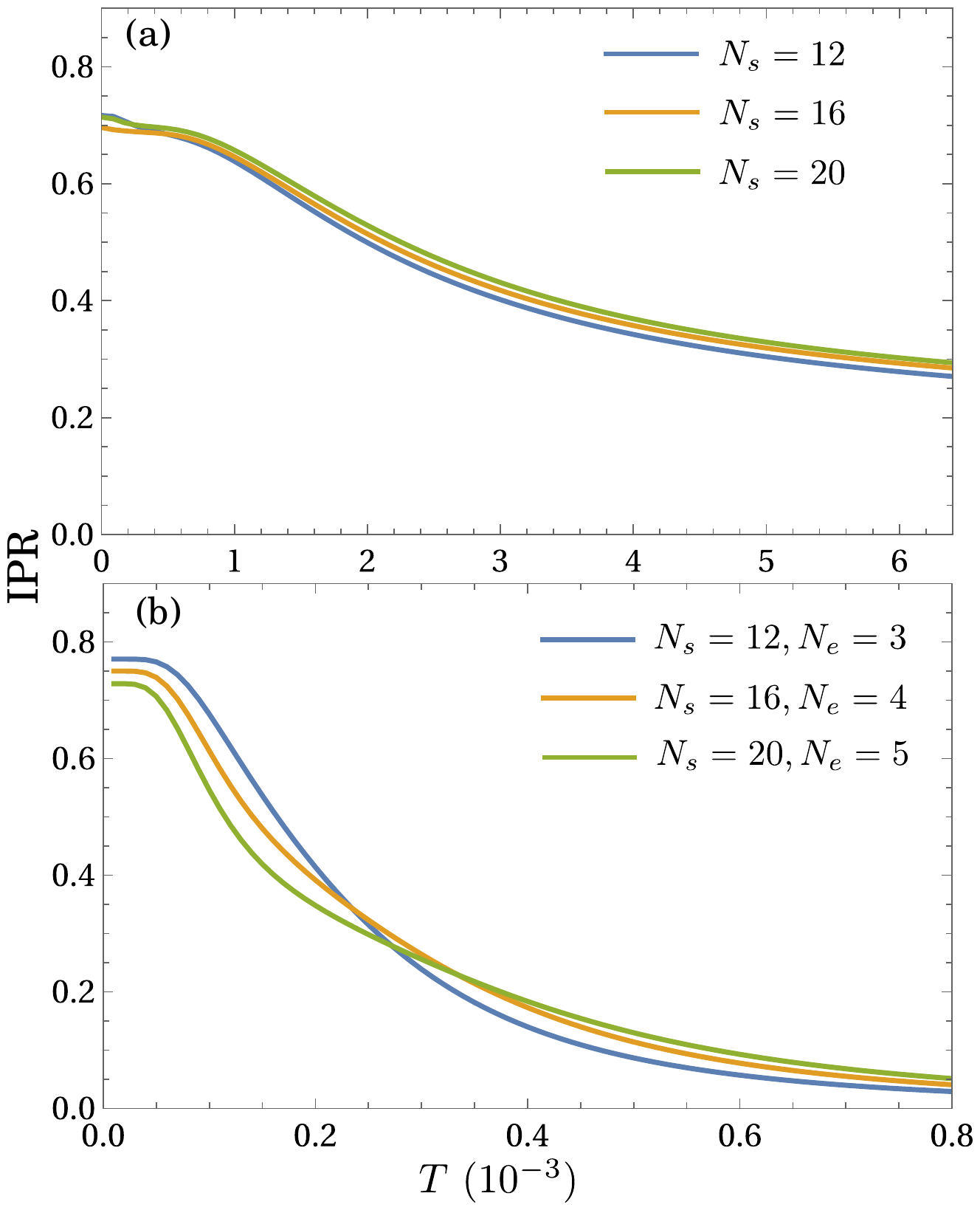}
	\caption{IPR as a function of temperature for different system size. (a) $N_{\text{impurity}}=1$, $N_e=4$, $1/a=0.02$ and $V=5$. (b) $N_{\text{impurity}}=1$, $1/a=0.01$ and $V=0.2/16\times N_s$. \label{fig4}}
   \end{figure}
		
    \begin{figure}
			\includegraphics[width=0.4\textwidth]{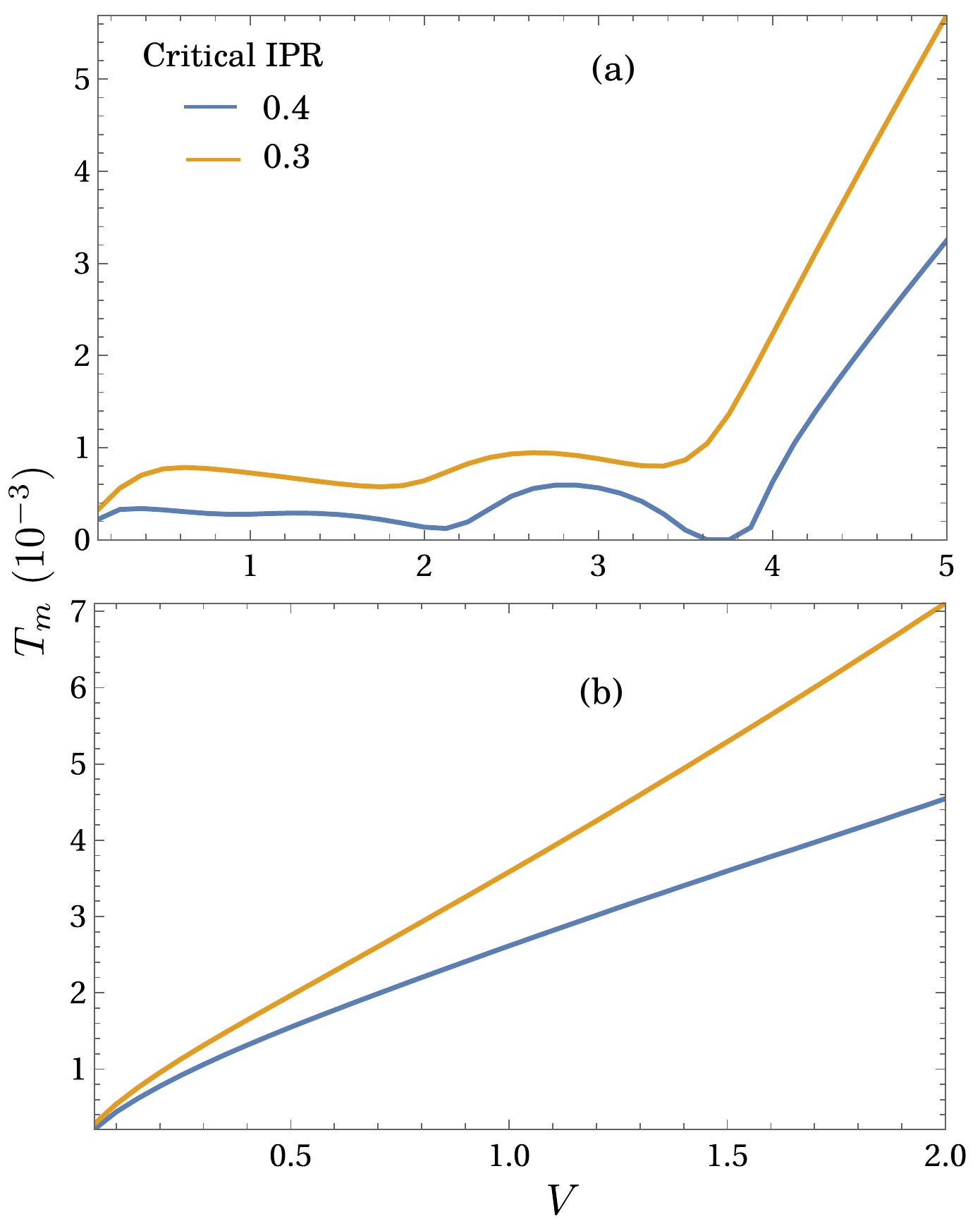}
			\caption{Melting temperature as a function of disorder strength under two IPR criteria with $1/a=0.02$. (a) Single impurity $X=1$. (b) Two impurities $X_1=1, X_2=9$. For the former, the transitional $V\sim 4$, while the critical $V$ is significantly reduced as in the latter two impurity-case. These trends manifest consistently under different critical IPR.  \label{fig5}}
    \end{figure}
   
 In this section, we study how the Wigner-Anderson interplay is described under different criteria. We first demonstrate the melting of the localized states (either Wigner crystal or Anderson insulator) at different system sizes. There are three extensive numbers in our simulation, i.e. the number of impurities, electrons and lattice sites. The conventional scaling analysis requires all these numbers to scale proportionally, i.e. if we start with $N_{\text{impurity}}=1$, $N_e=4$ and $N_s=16$ the next step involves $N_\text{impurity}=2$, $N_e=8$ and $N_s=32$. As such, the system size quickly grows beyond the capability of the exact diagonalization. However, at the limit of large and small $V$, we can approximate the scaling as follows. 
   
   For large $V$, the localized state at low temperature is mostly induced by the binding impurity potential and uncorrelated. Accordingly, the most important factor is the ratio, $N_e/N_{\text{impurity}}$ while $N_e/N_s$ is much less relevant. As a result, we only increase $N_s$ while keeping  the number of electrons and impurities unchanged (the electron-electron interaction is still included). Consistent with the uncorrelated nature of state, the insensitivity of the IPR with the system size clearly manifests in Fig.~\ref{fig4}(a). We expect this insensitivity to exist as well if we glue multiple identical lattices together instead of just increases the number of lattice sites. So the liquid-Anderson insulator transition is stable in the thermodynamic limit.
   
 On the contrary, for small $V$, the low-temperature localized state is a Wigner crystal so the electron filling $N_e/N_s$ is important here, while the impurity potential only pins down the crystal. As a result, we scale $N_e$ and $N_s$ proportionally, while keeping the $N_\text{impurity}$ fixed at $1$ but modifying the strength $V\to V' = V\times L'/L$. Figure~\ref{fig4}(b) also shows a moderate stability of the IPR with respect to system size. Therefore, we can justify the use of scaled IPR as a criterion to separate  the localized and extended phase for both the Anderson insulator and Wigner crystal. We note that the modulation of IPR with temperature is a crossover in our simulation and there is indeed some degree of ambiguity in choosing the threshold IPR. However, as shown in Fig.~\ref{fig5}, regardless of the value of critical IPR, for the single-impurity case, the observed localized phase is an Anderson insulator (Wigner crystal) for $V>4~(<4)$ as characterized by the linear ramp of $T_m$ with respect to $V$; while for the two-impurity case, this critical $V$ is significantly reduced.